\documentclass[prb,preprint,showpacs,letterpaper]{revtex4}

\usepackage{graphicx} 
\usepackage{bm}
\usepackage{float}
\usepackage{amsmath,amssymb,dcolumn,setspace}

\begin{document}

\title{Transport Properties of Carbon Nanotube C$_{60}$ Peapods}

\author{C. H. L. Quay}

\author{John Cumings}
\altaffiliation[Present address ] {Department of Materials Science \& Engineering, University of Maryland, College Park, MD 20742-2115.}
\affiliation {Physics Department, Stanford University, Stanford, CA 94305-4060, USA.}

\author{S. J. Gamble}
\affiliation{Applied Physics Department, Stanford University, Stanford, CA 94305-4090, USA.}

\author{A. Yazdani}
\affiliation{Department of Physics, Princeton University, Princeton, NJ 08544, USA.}

\author{H. Kataura}
\affiliation{Nanotechnology Research Institute, National Institute of Advanced Industrial Science and Technology, Central 4, Higashi 1-1-1, Tsukuba, Ibaraki 305-8562, Japan.}

\author{D. Goldhaber-Gordon}
\affiliation{Physics Department, Stanford University, Stanford, CA 94305-4060, USA.}


\begin{abstract}
We measure the conductance of carbon nanotube peapods from room temperature down to 250mK. Our devices show both metallic and semiconducting behavior at room temperature. At the lowest temperatures, we observe single electron effects. Our results suggest that the encapsulated C$_{60}$ molecules do not introduce substantial backscattering for electrons near the Fermi level. This is remarkable given that previous tunneling spectroscopy measurements show that encapsulated C$_{60}$ strongly modifies the electronic structure of a nanotube away from the Fermi level.~\cite{hornbaker}
\end{abstract}


\maketitle

In the fifteen years since their discovery,~\cite{ijima} carbon nanotubes' electronic properties have generated considerable excitement in the physics and engineering communities. In addition to being ideal one-dimensional electronic systems, carbon nanotubes hold promise for use as transistors,~\cite{javey} memory~\cite{radosavljevic,fuhrernanolett} and logic elements,~\cite{bachtold} and field emitters.~\cite{rinzler} In recent years, it has become possible to synthesize supra-molecular structures by inserting smaller molecules such as C$_{60}$ fullerenes into nanotubes to form `peapods'.~\cite{monthioux} Early experiments have shown that the inclusion of fullerenes modifies the electronic structure of a nanotube at energies far from the Fermi level,~\cite{hornbaker} and that a peapod's conductance can depend on the choice of encapsulated species,~\cite{chiu:apa,shimada:pe} raising the prospect of novel transport phenomena in these molecules.

In this Letter we report measurements of the conductance of carbon nanotube peapods at temperatures from 250mK to room temperature. We were surprised to find that the addition of C$_{60}$ molecules does not significantly modify the transport (low energy) properties of nanotubes --- our devices exhibit a range of behavior similar to that previously seen in empty nanotubes. At room temperature, the nanotubes are semi-conducting or metallic; at low temperature, we observe Coulomb blockade, and both spin-1/2 and spin-1 Kondo effects. Here we discuss the overall behavior of our ensemble of devices and make some statistical statements about them. A detailed description of the Kondo effects will be published separately. 

Our devices are carbon nanotube C$_{60}$ peapods contacted by palladium source and drain electrodes, 250nm to 500nm apart. The peapods lie on a 500nm or 1$\mu$m thick thermal oxide atop a highly-doped silicon substrate, which acts as the gate. The peapods are synthesized by the sublimation technique described in Ref.~\onlinecite{kataura}. They are then dispersed by sonication in chloroform or ortho-dicholorobenzene. The dispersion is deposited on the substrate and allowed to dry. We locate the peapods relative to pre-existing alignment marks using atomic force microscopy (AFM), and fabricate the electrodes using standard electron-beam lithography techniques. All nanotubes studied are 1.5-4nm in diameter according to AFM measurements.

\begin{figure}[htbp!]
	\includegraphics[0, 0][86mm,85.3mm]{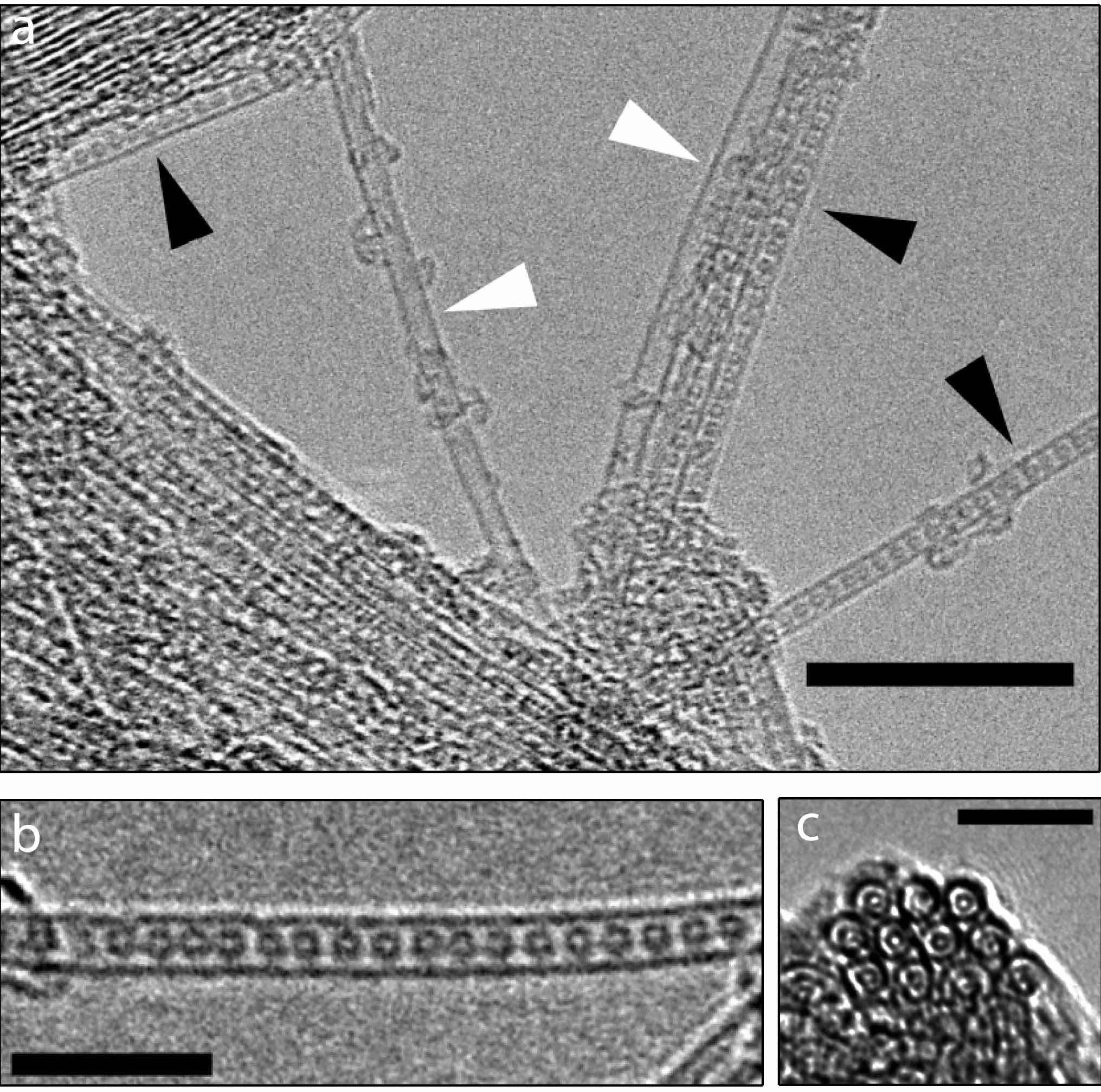}
	\caption{\label{fig:f1} (a) TEM image of bundles of carbon nanotubes deposited from our chloroform suspension, most filled with C$_{60}$. Arrows point to unobscured single nanotubes representative of those counted in our analysis --- black arrows to filled tubes, white ones to unfilled. The scalebar is 30nm long. (b) A single nanotube filled with C$_{60}$ peas. The scalebar is 5nm long. (c) A bundle of nanotubes viewed at an angle, showing the C$_{60}$ molecules inside. The scalebar is 5nm long.}
\end{figure}

Figure 1 shows representative transmission electron microscopy (TEM) images taken of nanotubes deposited from our dispersion. We deposited electrodes on 20 different nanotubes, of which 7 were found to be conductive at room temperature. The other 13 nanotubes are discounted from the analysis that follows as they are likely not connected due to handling or alignment problems.~\cite{footnote2}

Next, we perform a statistical analysis of our group of 7 nanotubes based on TEM images of our nanotubes deposited from the same ensemble. Such a statistical analysis is crucial, as no synthesis method yields 100\% filled peapods, and it is impractical to verify directly that a given nanotube in transport studies is filled with fullerenes --- TEM is the only established method for differentiating between filled and unfilled carbon nanotubes, and the specimen requirements for TEM imaging are incompatible with the standard geometry of nanotube transistors. From images such as those in Figure~1, we identify 109 single nanotubes, which were unobscured and in focus. Of these, 92 nanotubes are filled, and 17 empty. We observe no partially-filled tubes.

\begin{figure}[htbp!]
	\includegraphics[0, 0][86mm,50.3mm]{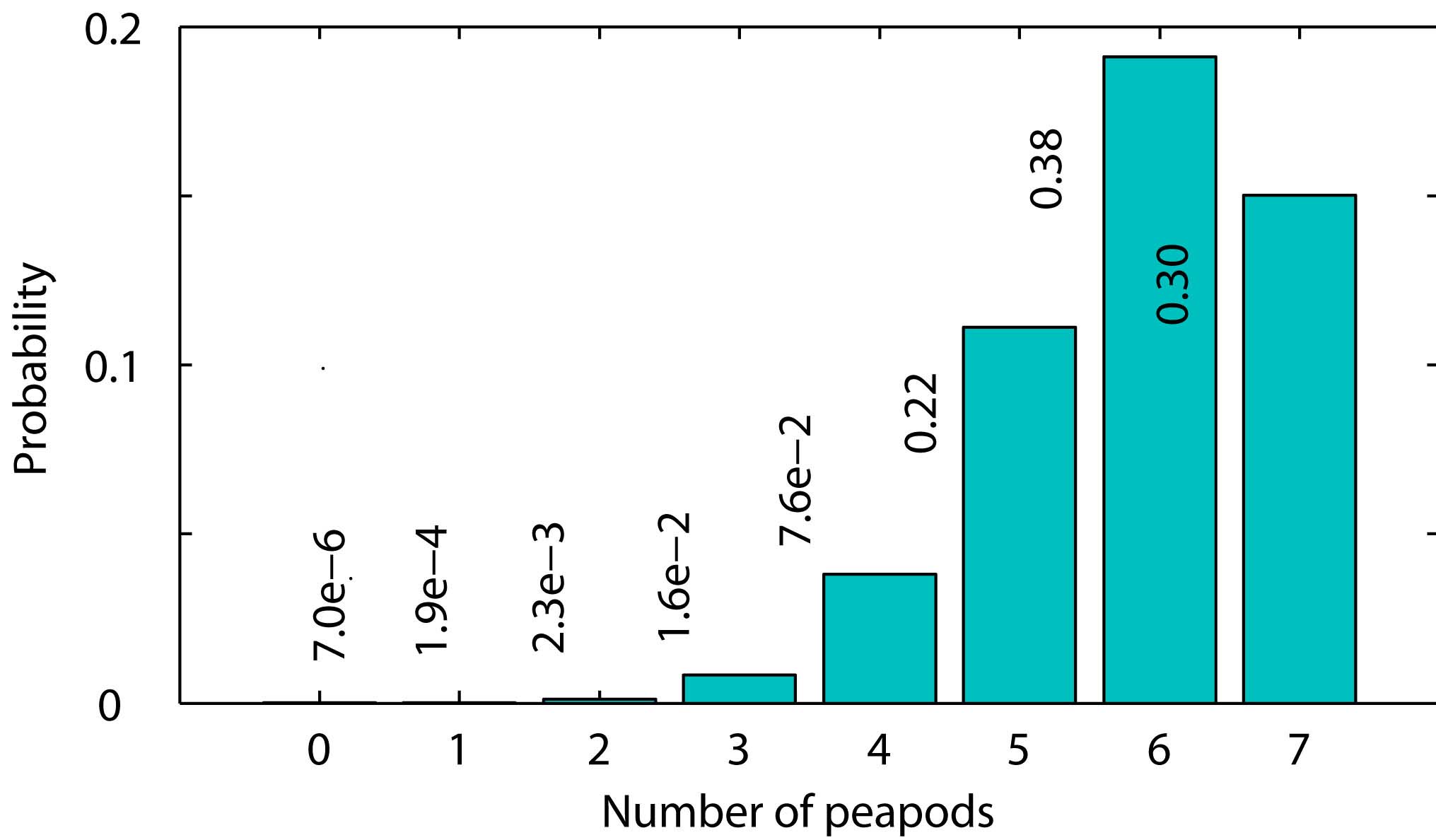}
	\caption{\label{barchart} Calculated probabilities, using Bayes's Theorem, for the number of filled nanotubes in our sample of 7, taking into account the proportion of filled nanotubes in our TEM images (92 out of 109).}
\end{figure}

Taking into account the frequency of filling in the nanotubes examined by TEM and assuming no prior knowledge of the fraction of filled tubes, we use Bayes's Theorem for continuous probability distributions~\cite{bayes, papoulis} to evaluate the probability that our 7 measured nanotubes included any specific number of filled tubes from 0 to 7. This information is presented in Figure~2. The expected number of filled tubes is found by this method to be 5.86. Details of these calculations are included in the Supplementary Material; however, we would like to emphasize here that our approach is more conservative than simply taking 92/109 as the fraction of filled nanotubes, yielding a higher probability that many of our nanotubes are unfilled. 

\begin{figure}[htbp!]
	\includegraphics[0, 0][86mm,90.6mm]{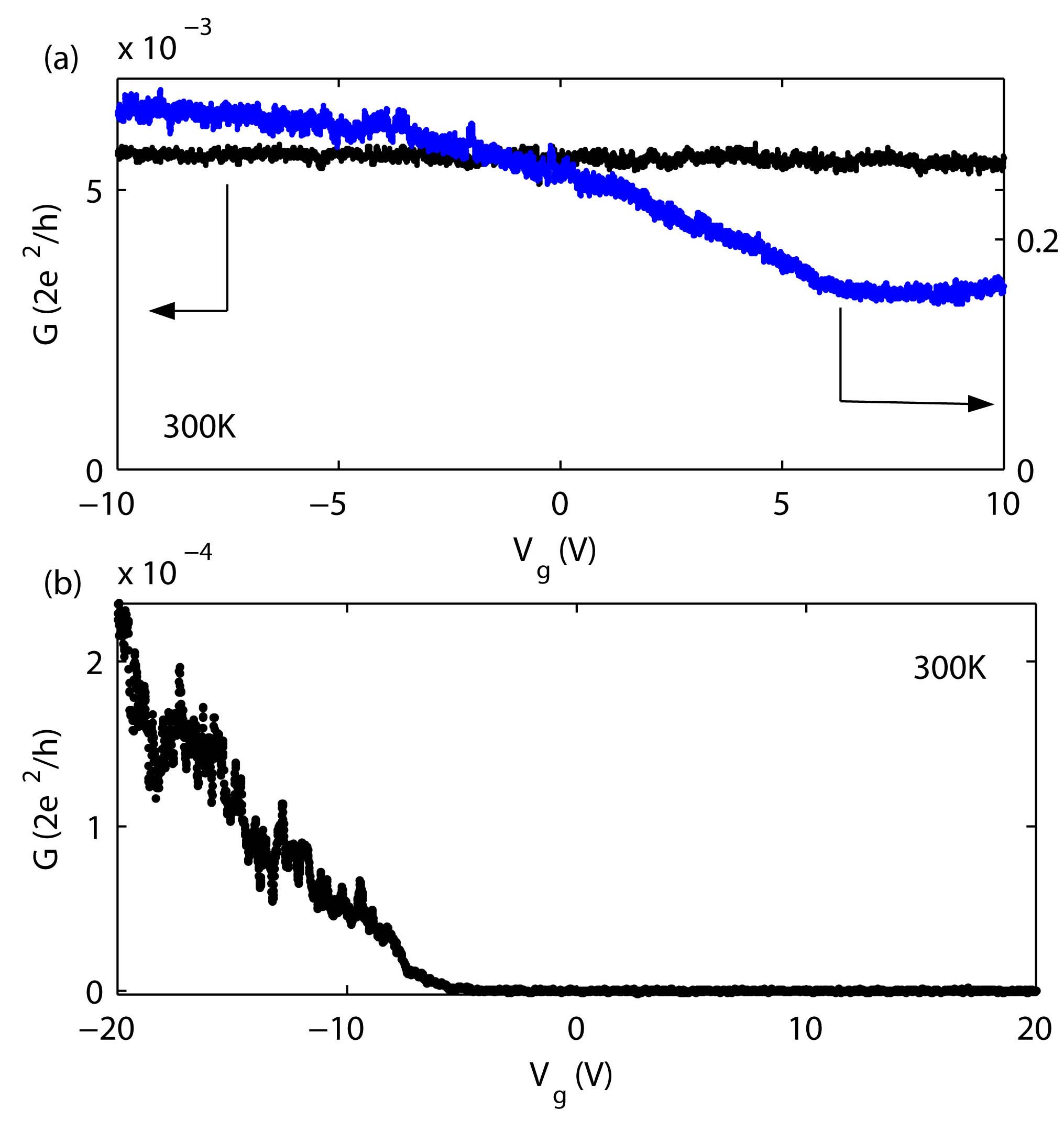}
	\caption{\label{f2} Our devices, which include some peapods (see Figure 2) show a range of room-temperature transport properties indistinguishable from those of unfilled nanotubes. (a) Room temperature linear conductance traces for devices exhibiting some (right axis) and no (left axis) change as the gate voltage is swept. (b) Room temperature linear conductance of a completely depletable semiconducting device.}
\end{figure}

Returning to the transport properties, Figures~3 and~4 show representative measurements of the conductance of our devices as a function of gate voltage at room temperature and at 250mK. 

In the room temperature measurements, we observe devices with conductances significantly modified by the gate as well as ones that are unaffected by it (Figure 3a): `semiconducting' and `metallic', respectively, in the conventional description of carbon nanotubes.~\cite{hamada,nygard:apa} Only a few devices with rather low overall conductances are completely depletable (Figure 3b). 

\begin{figure*}[htbp!]
	\includegraphics[0, 0][178mm,100.5mm]{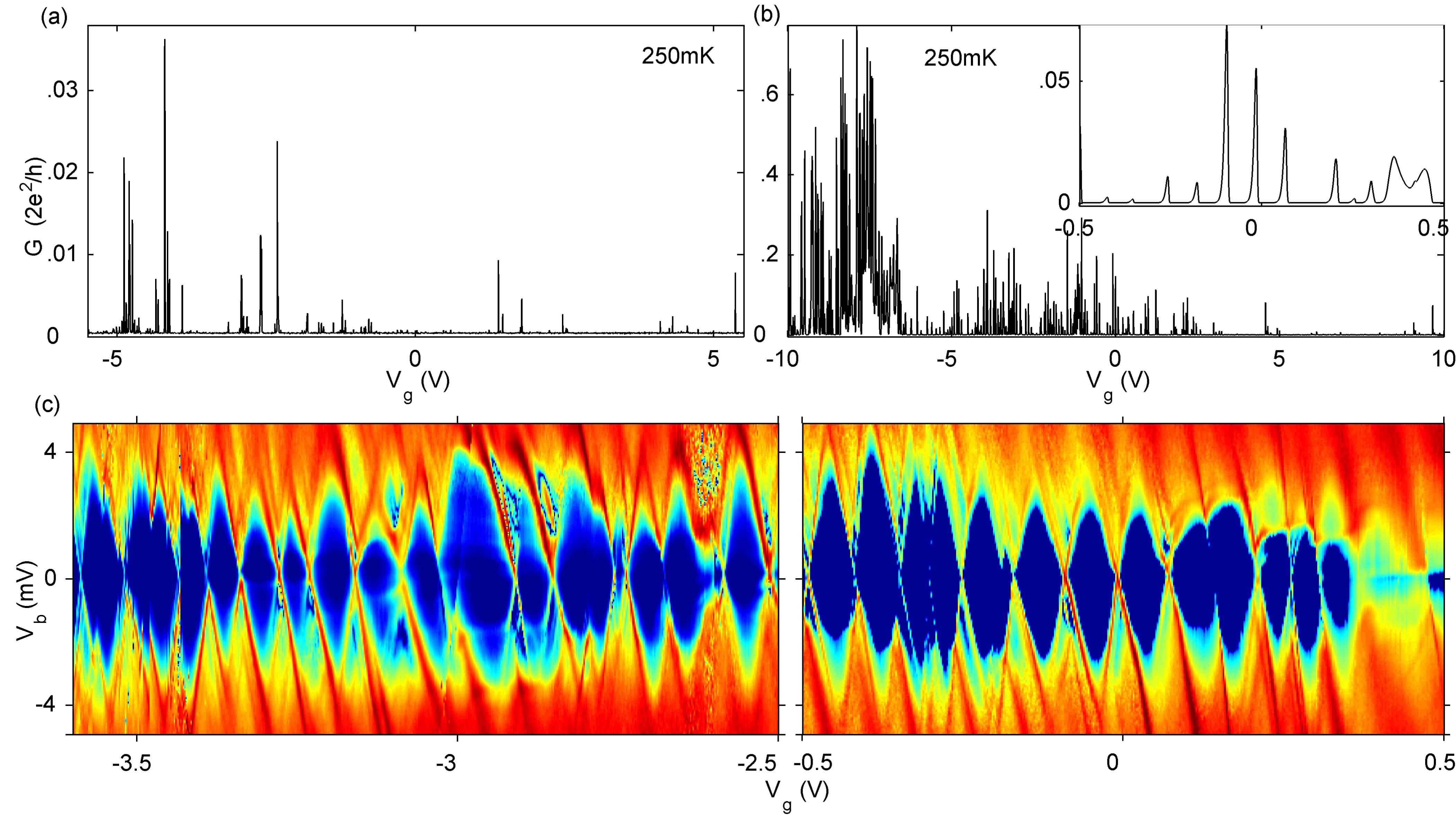}
	\caption{\label{f3} (a) At 250mK, device showing Coulomb Blockade. (b) Also at 250mK, device with higher conductance showing Coulomb Blockade. (Inset) Detail showing regularity of peaks, which continues over the whole range. (c) Conductance versus bias and gate voltage of device in (b). The color scale is blue (low) to red (high conductance). Regular diamonds indicate that this is a single quantum dot.}
\end{figure*}

At 250mK-350mK, two of the seven nanotubes in our ensemble have undetectably low conductance. The question naturally arises as to whether these, and only these, are peapods. As seen in Figure~2, we find that the probability that three or more of our tubes in this sample of seven are filled is 99.75\%. (See Supplementary Material for details.) It is therefore practically a certainty that one or more of our single quantum dot devices are formed on peapods. 

All devices measurable at low temperature show Coulomb Blockade behavior, but with widely varying peak conductances. Representative traces are shown in Figure~4. In most devices, measurements of conductance versus gate and bias voltages show Coulomb diamonds (Figure 4c is representative) indicating that each device acts as a single quantum dot.~\cite{footnote1} The charging energies seen (5-40meV) are consistent with quantum dots formed by tunnel barriers at the contact electrodes, indicating that electrons are delocalized over the 250-500nm length of nanotube between contacts. If, as argued above, some of the devices are formed from peapods, we may conclude that the encapsulated C$_{60}$ does not introduce substantial backscattering of electrons passing through the nanotube. This somewhat surprising result is consistent with data from recent photoemission studies.~\cite{shiozawa} The absence of backscattering in peapods may be due to the long wavelength of the perturbation introduced by the encapsulated C$_{60}$. Due to an unusual bandstructure, backscattering in single-walled carbon nanotubes is expected to require a very large momentum transfer, which can only be produced by a nearly atomically sharp perturbation or a perturbation so large that it locally depletes the tube.

To our knowledge, there has been only one previous report of transport measurements on a nanotube believed to contain C$_{60}$ molecules~\cite{yu} (though several studies have been published on metallofullerene peapods). In these measurements, Yu et al. found modulation of conductance by gate and bias voltages on large energy scales, and weak conductance at zero bias, suggesting formation of multiple dots in series within the nanotube. 

In conclusion, we have measured the transport properties of carbon nanotube samples including some C$_{60}$ peapods at room temperature and at 250-350mK, and have done the first careful statistical analysis of such an assembly of devices. Our results indicate that C$_{60}$ peapods do not differ collectively from nanotubes in their electronic transport characteristics. We note that this corroborates earlier STM work,~\cite{hornbaker} where C$_60$ peas were found to induce significant perturbations in electronic structure of a nanotube only at much higher energies than were accessed in our present measurements.

Photoemission studies nevertheless suggest that other peapod species may yield more exotic behavior in transport --- for example, a Tomonaga-Luttinger- to Fermi-liquid transition with increased potassium doping.~\cite{rauf} A more detailed picture of the range of transport properties of peapods may emerge when transport measurements can be combined with \textit{in situ} structural characterization. Meyer et al.~\cite{meyer} have commenced work in this direction.

We thank Rafael de Picciotto for technical assistance and scientific advice, and acknowledge technical assistance from Yi Qi. This work was performed in part at the Stanford Nanofabrication Facility of NNIN, and supported by U.S. Air Force Grants No. FA9550-04-1-0384 and F49620-02-1-0383. CQHL acknowledges support from Gabilan Stanford Graduate and Harvey Fellowships, and DGG Fellowships from the Packard and Sloan Foundations.

\appendix*

\section{Bayes's Theorem \label{append} }

Bayes's Theorem for continuous probability distributions~\cite{bayes, papoulis} states \begin{equation} \label{contdist}
f(x|y) = \frac{f(y|x)f(x)}{\int\limits_{-\infty}^{\infty} f(y|x) f(x) \,dx}.
\end{equation}

Here $f(x)$ is the marginal density of the random variable \textbf{x}, and $f(y|x)$ is the conditional density of the random variable \textbf{y} given \textbf{x}$= x$. For us, \textbf{x} is the fraction of filled nanotubes in our ensemble and \textbf{y} the probability of finding 92 filled tubes in a sample of 109 drawn from that ensemble.

Equation \eqref{contdist} thus gives us $f(x|y)$, the posterior distribution of \textbf{x} given that we found 92 filled nanotubes out of 109. $f(y|x)$ is simply the binomial distribution given any particular \textbf{x}$= x$, \begin{equation} \label{drawprob}
f(y|x) = \binom{109}{17}x^{92}(1-x)^{17}.
\end{equation}

To guard against over-estimating the number of filled nanotubes in our sample, we assume a uniform prior distribution of \textbf{x}, i.e. $f(x) = 1$, making our calculations more conservative than if we had simply used `92/109' as the fraction of filled nanotubes in our ensemble. 

As equation~\ref{contdist} gives a `probability distribution', the conditional expected value for any function, $g(x)$ is \begin{equation} \label{expectation}
\bar{g} = \int\limits_{-\infty}^{\infty} f(x|y) g(x)\,dx,
\end{equation} i.e. $g(x)$ multiplied the `probability' of each $x$ and integrated over all $x$.

For example, for a given value $x$ of the random variable \textbf{x}, by definition the probability that any particular nanotube in the ensemble is filled is $g(x) = x$. Putting this into equation~\ref{expectation}, which accounts for the earlier `92 out of 109' observation, we find that the likelihood that any randomly-chosen nanotube is filled is $\bar{g} = 31/37$. Thus, the expected number of filled tubes in our sample of 7 is 5.86.

Similarly, to obtain the probability that a specific number $n$ of our 7 nanotubes are filled, we substitute \begin{equation} \label{chart}
g(x) = \binom{7}{n}x^{n}(1-x)^{7-n}
\end{equation} into equation~\ref{expectation} to produce Figure 2.


\begin{thebibliography}{21}
\bibitem[{foo({\natexlab{a}})}]{footnote2}
\bibinfo{note}{We deposited multiple electrodes on each of 20 nanotubes. All
  but one of the electrically-inactive nanotubes were from a batch with known
  handling issues. Even at 0.5V bias and $\pm20V$ on the gate at room
  temperature, every adjacent electrode pair on these nanotubes showed no
  discernible current above a $\sim4pA$ noise floor, implying a resistance
  above 100G$\Omega$. In contrast, even the most resistive of the `conductive'
  nanotubes passed several nA of current under similar conditions,
  corresponding to a nanotube resistance of $\lesssim200M\Omega$.}
\end{thebibliography}

\end{document}